**Discovery of a magnetic topological semimetal $Sr_{1-y}Mn_{1-z}Sb_2$ ($y$, $z$ < 0.10)**

J.Y. Liu [1†], J. Hu [1†], Q. Zhang[2†], D. Graf [3], H.B. Cao[4], S.M.A. Radmanesh[5], D.J. Adams [5], Y.L. Zhu[1], G.F. Cheng [1], X. Liu[1], W. A. Phelan[2], J. Wei[1], D. A. Tennant[4], J. F. DiTusa[2], I. Chiorescu [3,6], L. Spinu[5] and Z.Q. Mao[1*]

[1] Department of Physics and Engineering Physics, Tulane University, New Orleans, LA 70018

[2] Department of Physics and Astronomy, Louisiana State University, Baton Rouge, LA 70803

[3] National High Magnetic Field Lab, Tallahassee, FL 32310

[4] Quantum Condensed Matter Division, Oak Ridge National Laboratory, TN 37830

[5] Department of Physics and Advanced Materials Research Institute, University of New Orleans, New Orleans, LA 70148

[6] Department of Physics, Florida State University, Tallahassee, FL 32306

[†]These authors contribute equally to this work

[*]E-mail: zmao@tulane.edu

**Recent discoveries of topological Weyl semimetals (WSM) in noncentrosymmetric monopnictides TX (T=Ta/Nb, X=As/P) [1-7] and photonic crystals [8] have generated immense interests since they represent new topological states of quantum matter. WSMs evolve from Dirac semimetals (DSMs) in the presence of the breaking of time reversal symmetry (TRS) or space inversion symmetry [1,2]. The WSM phases in TX and photonic crystals are due to the loss of space inversion symmetry. For TRS breaking WSMs, despite numerous theoretical and experimental efforts [9-11], only one example, $YbMnBi_2$, has recently been**

**reported [12] and its TRS breaking is predicted to be caused by a net ferromagnetic (FM) component of a canted antiferromagnetic state (CAFM) [12]. In this letter, we report a new type of magnetic topological semimetal phase arising from 2D Sb layers in $Sr_{1-y}Mn_{1-z}Sb_2$ ($y$, $z < 0.1$) whose relativistic fermion behavior, including a π Berry phase and high carrier mobility, was revealed from quantum transport measurements. Neutron scattering studies show this material is in a FM state for 304 K < T < 565 K, which evolves into a CAFM state with a FM component for T < 304 K. The combination of relativistic fermion behavior and ferromagnetism make this material a promising candidate for exploring the long-sought magnetic WSM.**

3D Dirac semimetals (DSMs) can be viewed as a 3D analogue of graphene and are characterized by linear energy-momentum dispersion near the Fermi level along all three momentum directions[13-18]. The linear band crossing point, i.e. the Dirac point, is protected against gap formation by crystal symmetry. Such unique band structures of 3D DSMs result in peculiar exotic properties such as high bulk carrier mobility [19], large linear magnetoresistance[19-21], and quantum spin Hall effect[13]. The experimentally established examples of 3D DSMs include $Na_3Bi$ [14], $Cd_3As_2$ [16-18], TlBiSSe [22], $ZrTe_5$ [23] and ZrSiS [24].

DSMs can be regarded as parent materials of Weyl semimetals (WSMs). When either time-reversal symmetry (TRS) or space inversion symmetry is broken, DSMs evolve into WSMs[1,2]. That is, each Dirac point characterized by chiral symmetry splits into two Weyl points with opposite chirality, through which energy bands disperse linearly. Each Weyl point can be seen as a magnetic monopole in momentum space. Remarkable characteristics of Weyl state include surface Fermi arcs connecting Weyl points [9] and the chiral anomaly, which originates from charge pumping between Weyl points with opposite chirality and is manifested as negative longitudinal magnetoresistance (LMR) [25-27]. Among current 3D DSMs, the chiral anomaly effect due to the field-induced TRS breaking have been observed in $Na_3Bi$ [28], $ZrTe_5$ [23] and topological insulators such as $Bi_{1-x}Sb_x$ [29]. Recently, a Weyl state due to spontaneous TRS breaking was also found in $YbMnBi_2$ [12]. WSMs generated by the broken space inversion symmetry have been experimentally realized in noncentrosymmetric monopnictides TX (T=Ta/Nb, X=As/P)[3-7] and photonic crystals [8], Both chiral anomaly effect [30-34] and surface Fermi arc have been experimentally observed in TX [3-7].

In this letter, we report a discovery of a new type of quasi-two-dimensional (2D) topological semimetal phase arising from 2D Sb layers in $Sr_{1-y}Mn_{1-z}Sb_2$ ($y$, $z < 0.1$). A distinct aspect of this material lies in its ferromagnetic properties. Neutron scattering measurements reveal a FM transition at 565K, followed by a transition to a canted antiferromagnetic state (CAFM) with a net ferromagnetic (FM) component, similar to the predicted CAFM state for $YbMnBi_2$ [12]. The magnitude of the FM component, which strongly depends on Sr and Mn non-stoichiometry, was found to be coupled to quantum transport properties, implying the effect of TRS breaking on the electronic band structure. These findings make $Sr_{1-y}Mn_{1-z}Sb_2$ a promising candidate for exploring a novel magnetic Weyl state.

$SrMnSb_2$ is a close cousin of $SrMnBi_2$ and $CaMnBi_2$, which have been reported to be quasi-2D Dirac materials [35-37]. The 2D Bi square nets harbor Dirac Fermions in both materials and the Dirac cones are anisotropic, with a small gap opening at the Dirac point due to strong spin-orbital coupling [35,37]. Several other relevant compounds, including $LaAgBi_2$[38] and $LaAgSb_2$[39], have also been suggested to be Dirac materials. However, for $SrMnSb_2$, the material studied in this work, the early first-principle calculations [40] suggested it is not a Dirac material due to the orthorhombic distortion on the Sb layers which leads Sb atoms to form zig-zag chains as shown in Fig. 1a, but Dirac fermions are expected to be present in this material under pressure [40]. Our discovery of a magnetic topological semimetal phase in $Sr_{1-y}Mn_{1-z}Sb_2$ is surprising and the nonstoichiometric composition of our samples may account for our experimental results.

Our $Sr_{1-y}Mn_{1-z}Sb_2$ ($y$, $z < 0.1$) single crystals were prepared using self-flux method (see the method section for the details of the synthesis). The crystals are plate-like, with the biggest piece reaching dimensions of ~7×6×1 $mm^3$ (see the inset to Fig. 1b). The single-crystal neutron scattering measurements reveal that the synthesized material crystalizes in an orthorhombic structure with the space group *Pnma* despite Sr and Mn nontoichiometry, similar to the previously-reported structure on stoichiometric $SrMnSb_2$ [41]; no structural transition was observed down to 5K. The lattice and other structural parameters obtained from the neutron diffraction refinement are presented in the Supplementary Information (SI). Figure 1b shows the XRD pattern of the (*h*00) plane for a typical crystal. The sharp diffraction peaks indicate high crystalline quality. The compositions of our single crystals were analyzed using an energy dispersive X-ray spectrometer, which shows that the actual composition involves Mn and Sr deficiency as described by $Sr_{1-y}Mn_{1-z}Sb_2$ ($y$, $z < 0.1$). Interestingly, the Mn or Sr nonstoichiometry was found to have a strong effect on the magnetic properties of $Sr_{1-y}Mn_{1-z}Sb_2$. Samples with larger Sr deficiency ($y$~0.08, $z$ ~ 0.02) is accompanied by stronger FM behavior, while weaker FM behavior occurs in samples with enhanced Mn deficiency ($y$ ~ 0.01-0.04. $z$ ~ 0.04 -0.1). According to the magnitude of FM saturated moment $M_s$, we categorize our samples into three types; $M_s$ ~ 0.2-0.6 $\mu_B$/Mn for type A, 0.04-0.06 $\mu_B$/Mn for type B and 0.004-0.006 $\mu_B$/Mn for type C. Detailed comparisons of magnetic properties between type A, B and C samples will be given below.

We will first show electronic transport properties of Type A samples and then compare them with those of type B and C samples. All the transport data presented in Fig. 1and 2 were collected on type A samples. As seen in Fig. 1c, both in-plane ($\rho_{xx}$) and out-of-plane ($\rho_{out}$) resistivity exhibit metallic temperature dependences. The in-plane residual resistivity $\rho_{xx,0}$ is ~ 27 μΩ cm, and the residual resistivity ratio $\rho_{xx}$(300K)/$\rho_{xx}$(2K) ~ 29. The $\rho_{out}/\rho_{xx}$ ratio increases markedly with decreasing temperature, reaching 609 at 2K. Such a large $\rho_{out}/\rho_{xx}$ ratio suggests quasi-2D electronic band structure. We conducted magnetotransport measurements along both in-plane and out-of-plane directions for type A samples. As shown in Fig. 1d-1f, the magnetoresistivity MR (= $[\rho(B)-\rho(0)]/\rho(0)$) along both directions exhibit strong Shubnikov-de Hass (SdH) oscillations for $T$<30 K. The field was applied parallel to the out-of-plane direction for both measurements. The oscillations extend to low field range (see Fig. 1e) and the relative oscillation amplitude $\Delta\rho_{out}/\rho_{avg}$ can reach 100% near 23T at 1.6 K. Such SdH oscillations are much stronger than those observed in $SrMnBi_2$ [35,42]. Strong SdH oscillations imply high carrier mobility, which is confirmed in our Hall effect measurements as shown below.

We found significant signatures of relativistic fermions from the analyses of SdH oscillations. First, both in-plane and out-of-plane MR exhibit SdH oscillations with a small frequency, which can be easily identified from the plots of oscillatory component $\Delta\rho_{xx}$ (or $\Delta\rho_{out}$) vs. $1/B$ (see Figure S2 in SI). The oscillation frequencies are ~ 69 T and 67 T, respectively, for $\Delta\rho_{xx}(B)$ and $\Delta\rho_{out}(B)$. Such a small oscillation frequency was verified in multiple samples. From the SdH oscillation frequency $F$, the extremal cross-sectional area $A_F$ of the Fermi surface (FS) can be obtained using the Onsager relation $F=(\Phi_0/2\pi^2)A$. The frequency of 67 T corresponds to $A_F$ = 0.64(0) $nm^{-2}$, about one half of $A_F$ (=1.45$nm^{-2}$) probed in the same field configuration for $SrMnBi_2$ [35]. Such a small value of $A_F$ indicates a small FS. Moreover, we also measured the dependence of $F$ on the magnetic field orientation angle $\theta$ (*i.e.* the angle between the out-of-plane and the magnetic field). As we show in SI, the measured $F(\theta)$ can be fitted with a $F_0/\cos\theta$ function ($F_0$, the oscillation frequency for the field along the out-of-plane direction), suggesting the FS responsible for the SdH oscillations in $Sr_{1-y}Mn_{1-z}Sb_2$ is quasi-2D, consistent with the aforementioned quasi-2D transport properties.

Secondly, the effective quasiparticle mass $m^*$ extracted from the temperature damping of SdH oscillation amplitude is found to be small. We present the Fast Fourier Transform (FFT) spectra of $\Delta\rho_{out}(B)$ for various temperatures in Fig. 2a and the temperature dependence of normalized FFT amplitude in the inset. In general, $m^*$ can be obtained from the fit of the temperature dependence of FFT amplitude to the Lifshitz-Kosevich (LK) equation, *i.e.* $\Delta\rho/\rho_0 \propto \alpha T/\sinh(\alpha T)$ where $\alpha = (2\pi^2 k_B m^*)/(\hbar eB)$ and $\rho_0$ is the zero field resistance. However, as seen in the inset of Fig. 2a, the LK formula barely fits our data in the full temperature range of 2-50 K due to the steep increase of the oscillation amplitude below 7 K [see the black curve with poor fitting, which yields $m^* \sim 0.22m_0$ ($m_0$, the free electron mass)], but the best LK fit can be achieved in the 7-50 K range (the red fitting curve), which yields $m^* \sim 0.14m_0$. Similar phenomena also occur to the LK fit to the temperature damping of FFT amplitude of $\rho_{xx}$ and the $m^*$ obtained from the fit in 10-50 K range is $\sim 0.16m_0$ (see SI). The deviation from the LK fit below 10 K is unclear and deserves further investigation. The combination of small $m^*$ and small FS suggests $Sr_{1-y}Mn_{1-z}Sb_2$ likely harbors relativistic fermions.

To seek further evidence for relativistic fermions in $Sr_{1-y}Mn_{1-z}Sb_2$, we examined the Berry phase $\Phi_B$ accumulated along cyclotron orbits. For a Dirac/Weyl system, pseudo-spin rotation under a magnetic field should result in a non-trivial Berry phase. Therefore non-trivial Berry phase is generally considered to be a key evidence for topological semimetals. All current bulk Dirac (e.g. $Cd_3As_2$ [21,43] and $SrMnBi_2$ [35]) or monopnictide Weyl systems [30-34] are all found to exhibit non-trivial Berry phase. The Berry phase can be accessed from the fan diagram of Landau levels (LL) of SdH oscillations, *i.e.* the LL index $n$ plot as a function of the inverse of magnetic field $1/B$. For a 2D Dirac system, the intercept $\beta$ on the $n$-axis of the LL index plot is expected to be 1/2 [44], for which the corresponding Berry phase is $2\pi\beta = \pi$. The Weyl system should follow a similar scenario due to similar linear band dispersions. In Figs. 2b, we present the LL index plot for $Sr_{1-y}Mn_{1-z}Sb_2$ where $n$ and $1/B_n$ are defined from the minimum positions of conductivity $\sigma_{xx} = \rho_{xx}/(\rho_{xx}^2+\rho_{xy}^2)$ (where $\rho_{xx}$ and $\rho_{xy}$ are longitudinal and transverse (Hall) resistivity respectively), which is a commonly accepted practice for other relativistic fermion systems [45,46]. The inset of Fig. 2b illustrates the assignment of $n$ for $\Delta\sigma_{xx}$, the oscillatory component of conductivity $\sigma_{xx}$ which is derived from the $\rho_{xx}$ and $\rho_{xy}$ data displayed in Fig. 1f. More discussions on the definitions of $n$ and $1/B_n$ are given in SI. Clearly the data points in the LL index plot can be fitted to a linear dependence very well and the intercept $\beta$ on the $n$-axis derived from the linear extrapolation of the fitted line is ~ 0.46, in good agreement with the expected intercept of 1/2 for a 2D Dirac/Weyl system. This result clearly demonstrates a π Berry phase accumulated along cyclotron orbit, thus providing a key evidence for relativistic fermions in $Sr_{1-y}Mn_{1-z}Sb_2$.

Another distinct characteristic of relativistic fermions is high carrier mobility [19,47], which is indeed observed in the Hall effect measurements of $Sr_{1-y}Mn_{1-z}Sb_2$. As shown in Fig. 1f, $\rho_{xy}$ shows linear field dependence in low field range, consistent with a single band scenario. Like $\rho_{xx}$, $\rho_{xy}$ also exhibits strong SdH oscillations in high field range. The Hall coefficient $R_H$ extracted from $\rho_{xy}$ remains positive in the whole measured temperature range (see the inset to Fig. 2c). Its variation with temperature appears to suggest a multiple-band effect, but its change of magnitude with temperature is much less significant compared to other multiple band systems. These observations suggest that although this is a multiple band system as suggested by first principle calculations [40], its transport properties are likely dominated by a linearly dispersing band. This argument is in accordance with the observation of strong SdH oscillations with a single frequency. The positive sign of $R_H$ indicates the dominant band should be a hole-like band. We have estimated the carrier mobility using $\mu = R_H/\rho_{xx}$ and find $\mu_h$ increases sharply below 50 K, up to $1.25\times10^4$ cm$^2$V$^{-1}$S$^{-1}$ at 2K. The carrier density $n_{\mathrm{Hall}}$ estimated from $R_H$ at 2 K is 1.7(8) × $10^{19}$ cm$^{-3}$. Both $\mu_h$ and $n_{\mathrm{Hall}}$ revealed in our experiments are within the ranges generally expected for Dirac/Weyl materials. Additionally, like $SrMnBi_2$ [48], $Sr_{1-y}Mn_{1-z}Sb_2$ also displays valley polarized interlayer conduction (see SI). That is, the in-plane magnetic field rotation can modulate the contribution of each valley to the out-of-plane conduction, resulting in a peculiar anisotropy in the out-of-plane conductivity as a function of the in-plane field orientation angle, $\sigma_{\mathrm{out}}(\phi)$. Quantitative analyses of $\sigma_{\mathrm{out}}(\phi)$ further demonstrate that the electronic transport properties of $Sr_{1-y}Mn_{1-z}Sb_2$ are dominated by anisotropic Dirac bands and the FS is corrugated along the $k_z$ direction.

Since $Sr_{1-y}Mn_{1-z}Sb_2$ is isostructural to $SrMnBi_2$ and $CaMnBi_2$, both of which have been established as Dirac materials[35-37], our observations of relativistic fermion behavior, including small FS and $m^*$, high carrier mobility, π Berry phase, as well as the valley polarized interlayer conduction, appear to suggest that $Sr_{1-y}Mn_{1-z}Sb_2$ should be categorized as a Dirac system. However, given that $Sr_{1-y}Mn_{1-z}Sb_2$ shows FM behavior as mentioned above, the TRS breaking is likely to drive this material to a Weyl state. In the top panel of Fig. 3a, we show the isothermal magnetization data at 5 K for typical type A samples which exhibit significant FM-type polarization with $M_s \sim 0.2\text{-}0.6\mu_B$/Mn for the magnetic field along the out-of-plane direction. All transport data we discussed above (Fig. 1 and 2) were collected on type A samples with $M_s \sim 0.6\mu_B$ (*i.e.* A#1). We also measured temperature dependence of magnetization and found the FM polarization occurs even at 400 K (see Fig. S6 in SI). In order to understand the nature of such FM behavior, we performed single-crystal neutron diffraction measurements using a type A sample with $M_s \sim 0.2\mu_B$ at 5 K (which is only the sample we could find large enough for neutron diffraction experiments). We found $Sr_{1-y}Mn_{1-z}Sb_2$ exhibits complex magnetic states. First, a long-range FM order with Mn moments along the *b*-axis occurs below $T_C$=565 K (see Fig. 4b). Fig. 4a shows the temperature dependence of the FM scattering intensity that overlaps with the nuclear scattering intensity at the (200) reflection, from which a clear FM transition around 565K can be seen, consistent with the FM polarization behavior at 400K in the magnetization measurements ( Fig. S6). The calculated FM ordered Mn moment along the *b*-axis is ~ 0.49(8) $\mu_B$/Mn. Second, the FM order parameter shows an unusual decrease below 300K, coinciding with the presence of a strong AFM order below 304 K, as seen in the temperature dependence of the (001) magnetic peak intensity in Fig.4c. Both the magnetic and nuclear structures are determined from the refinement of the neutron diffraction spectrum collected at 5 K. A *C*-type AFM structure was found to best fit the data. With the reduced FM moment as evidenced by the partially reduced (200) intensity below 304 K, we conclude that the AFM state in type-A sample should be a canted AFM state (Fig. 4d). The nearest-neighbor Mn spins with moments (~ 3.77(9) $\mu_B$) along the *a* axis are antiparallel aligned whitin the *bc* plane and parallel aligned along the *a* axis. The canting lead to a FM component along the *b*-axis, with the size of moment of ~0.2 $\mu_B$, consistent with that obtained from the magnetization measurements. Details of the magnetic structural refinement are given in SI. It is worth pointing out that the canted C-type AFM state probed in $Sr_{1-y}Mn_{1-z}Sb_2$ is analogous to the canted AFM state theoretically predicted forYbMnBi$_2$ [12] whose FM component is believed to be responsible for the TRS breaking, though neutron scattering data for this material has not been available [12]. Given that $Sr_{1-y}Mn_{1-z}Sb_2$ shares a similar layered structure with $YbMnBi_2$ and exhibits relativistic fermion behavior as discussed above, it is reasonable to anticipate that the FM component in $Sr_{1-y}Mn_{1-z}Sb_2$ might result in a TRS breaking Weyl state. ARPES measurements are called for to verify such a possibility. As indicated above, one remarkable signature of Weyl state is chiral anomaly, which is manifested as negative LMR. For $Sr_{1-y}Mn_{1-z}Sb_2$, we have indeed observed negative LMR in some type A samples with the current along the out-of-plane direction (see SI for details). As shown in Figure S8, the field and field-orientation dependences of the LMR appear to be in line with the chiral anomaly expected for a Weyl state. Although $Sr_{1-y}Mn_{1-z}Sb_2$ shows FM properties, the observed negative LMR should not be associated with magnetic scattering, since the magntoconductance follows $B^2$ dependence even when the magnetization becomes saturated above 2T (Fig. S8c). However, to argue for a Weyl state, more systematic magnetoresistance measurements up to high fields on samples with the current along the in-plane directions and ARPES studies are necessary.

Although transport measurements could not give direct evidence of a Weyl state induced by ferromagnetism in $Sr_{1-y}Mn_{1-z}Sb_2$, our experiments have revealed that the ferromagnetism strongly couples with quantum oscillations. As summarized in the top panel of Fig. 3d, the relative SdH oscillation amplitude exhibits a remarkable enhancement with the increase of the FM saturated moment $M_S$ for both $\rho_{xx}$ and $\rho_{out}$, from ~10% for type C to ~25-45% for type B, and ~30-60% for type A samples. Coincidentally, the SdH oscillation frequency $F$ statistically decreases with increasing $M_S$, except for a large deviation of one data point collected on a type C sample (Fig. 3d, bottom panel). Such a varying trend of $F$ with $M_s$ was also probed in in de Haas-van Alphen (dHvA) oscillations (Fig. 3d) of magnetization. dHvA oscillations can be seen clearly in type C samples; however, for type A and B samples, dHvA oscillations are observable only when the data are zoomed in (see Fig. S6), which can be attributed to their larger FM background.

Next, we will discuss the possible effect of disorders on quantum oscillations in $Sr_{1-y}Mn_{1-z}Sb_2$. In general, disorders are expected to suppress the SdH and dHvA oscillations. Given that, one may wonder if the suppressed SdH oscillations in type C samples could be associated with disorders. Such a possibility can be excluded by examining the correlation between the residual resistivity ratio RRR (which characterizes the disorder level) and the relative SdH oscillation amplitude. As seen in Fig. 3e, the SdH oscillations in $Sr_{1-y}Mn_{1-z}Sb_2$ do not show any systematic dependence on RRR for both $\rho_{out}$ and $\rho_{xx}$ and strong SdH oscillations occurs even for samples with RRR ~ 2-3, implying that disorder is not the key factor in governing the quantum oscillations in this system. This can be understood as follows. $Sr_{1-y}Mn_{1-z}Sb_2$ consists of alternating conducting Sb layers and insulating MnSb and Sr layers[40]. In such a layered structure, the disorders induced by Sr and Mn vacancies would have strong effect on the out-of-plane conduction but weak effect on carrier transport within the Sb planes; this explains much stronger variation in the temperature dependence of $\rho_{out}$ than $\rho_{xx}$ (Fig. 3c). In this scenario, we can reasonably expect strong quantum oscillations can still occur to the $Sr_{1-y}Mn_{1-z}Sb_2$ samples with high levels of disorders, as long as the field is applied parallel to the out-of-plane direction where the cyclotron motions responsible for quantum oscillations in both $\rho_{out}$ than $\rho_{xx}$ are less affected by disorders away from the Sb conduction planes.

In general, higher quantum mobility leads to narrower width of LL, which is a prerequisite for observing strong SdH/dHvA oscillations. Our observations of enhanced SdH oscillations in the samples with stronger ferromagnetism suggest enhanced carrier mobility in those samples. It is known that high mobility is a signature of topological relativistic system [19], therefore the correlation of ferromagnetism and mobility in $Sr_{1-y}Mn_{1-z}Sb_2$ might be understood in terms of the change in band structure driven by ferromagnetism. Indeed, this has been proposed for $YbMnBi_2$ [12]; first principle calculations have shown that in this material the spin degeneracy can be lifted by a FM component arising from a canted AFM state, thus closing the spin-orbital coupling induced gap and leading to the formation of Weyl nodes. In $Sr_{1-y}Mn_{1-z}Sb_2$, our observation of the decrease of the quantum oscillation frequency $F$ caused by enhanced ferromagnetism appears to support the *gap-closing* scenario due to TRS breaking. As seen in Fig. 3d, $F$ decreases from ~72 T to ~ 67 T when $M_s$ increases by two orders of magnitude from type C to A samples, which corresponds to a ~ 7% shrinking of the cross-section area of FS. Therefore, although we are not sure whether a Weyl state is realized in $Sr_{1-y}Mn_{1-z}Sb_2$ at this point, the systematic decrease of quantum oscillation frequency as well as the coincident increase of quantum mobility with increasing $M_s$ is consistent with the role of the TRS-breaking in generating a Weyl state in $YbMnBi_2$.

In summary, we have demonstrated $Sr_{1-y}Mn_{1-z}Sb_2$ hosts relativistic fermions from quantum transport property measurements. Typical behavior of relativistic fermions, such as small Fermi surface, light effective mass, high carrier mobility and a $\pi$ Berry phase have been observed. Unlike the Dirac materials $SrMnBi_2$ and $CaMnBi_2$ which show Neel-type AFM order [49], $Sr_{1-y}Mn_{1-z}Sb_2$ exhibits ferromagnetism for 304 K < $T$ < 565 K and a canted C-type AFM state with a FM component below 304 K. Both quantum oscillation amplitude and frequency were found to be coupled with the ferromagnetism. Our findings establish $Sr_{1-y}Mn_{1-z}Sb_2$ as the first example of magnetic topological semimetal, which offers a wonderful opportunity to seek the long-sought magnetic Weyl state and investigate the coupling of Dirac/Weyl fermions with magnetism.

## Methods

### Single Crystal Preparation and Magnetotransport Measurements

The $Sr_{1-y}Mn_{1-z}Sb_2$ ($y$ or $z < 0.1$) single crystals were synthesized using a self-flux method with the stoichiometric mixture of Sr, Mn and Sb elements. The starting materials were put into a small alumina crucible and sealed in a quartz tube in Argon gas atmosphere. The tube was then heated to 1050 °C for 2 days, followed by a subsequently cooling down to 400 °C at a rate of 3 °C/h. The plate-like single crystals as large as a few millimeters can be obtained (Fig. 1b, inset). The composition and structure of these single crystals was checked using Energy-dispersive X-ray spectroscopy and X-ray diffraction measurements.

The magnetotransport measurements were performed with a four-probe method, using a 14T Physics Property Measurement System at AMRI, UNO, and the 31 T resistive magnets at National High Magnetic Field Laboratory (NHMFL) in Tallahassee.

Two relatively large $Sr_{1-y}Mn_{1-z}Sb_2$ ($y$ or $z < 0.1$) ($y$ or $z < 0.1$) crystals with the mass of ~ 17 mg were investigated by neutron diffraction experiments on four-circle single-crystal diffractometer HB-3A, at the High Flux Isotope Reactor (HFIR), Oak Ridge National Laboratory, USA. One crystal was measured at 315 K above $T_N$ with a wavelength of 1.003 Å from the bent Si-331 (no λ/2 contamination) [50] to study the crystal structure, whereas the other crystal was measured with a wavelength of 1.542 Å with ~1.4% λ/2 contamination (Si-220 monochromator in high resolution mode (bending 150)) [50] to investigate the structure and magnetic structure at 5 K as well as order parameter of nuclear (200) and magnetic (001) Bragg peaks. The data in a wide temperature region of $5 < T < 760$ K were collected using a Closed Cycle Refrigerators (CCR) available at HB3A. SARAH representational analysis program[51] was used to derive the symmetry allowed magnetic structures. All the neutron diffraction data were analyzed using Rietveld refinement program Fullprof *suite* [52].

**Acknowledgement**

The authors thank Prof. Congjun Wu at UCSD for helpful discussions. The work at Tulane University was supported by the NSF under Grant DMR-1205469 (support for personnel and materials) and Louisiana Board of Regents under grant LEQSF(2014-15)-ENH-TR-24 (support for equipment purchase). The neutron scattering work is supported by the U.S. Department of Energy under EPSCoR Grant No. DE-SC0012432 with additional support from the Louisiana Board of Regents. The work at UNO is supported by the NSF under the NSF EPSCoR Cooperative Agreement No. EPS-1003897 with additional support from the Louisiana Board of Regents. The work at FSU and at the National High Magnetic Field Laboratory, is supported by the NSF grant No. DMR-1206267, the NSF CooperativeAgreement No. DMR-1157490 and the State of Florida. The authors also acknowledge support from grant DOE DE-NA0001979.

**Author contributions**

J.Y.L., J.H. and Q.Z equally contributed to this work. The single crystals used in this study was synthesized by J.Y.L. The magnetotransport measurements in 14 T PPMS was carried out by J.Y.L., D.J.A., Z.Q.M. and L.S. The high field measurements at NHMFL were conducted by J.H., D.G., S.M.A.R., I.C., L.S. and Z.Q.M. G.F.C., X.L., J.W. and W. A. P. contributed to x-ray structure characterization and crystal quality examination. J.H., J.Y. L. and Y.L.Z. performed magnetization measurements. Q. Z., H.B. C. J.F. D. and D.A. T. conducted neutron scattering experiments and analyses. J.Y.L., J.H., Y.L.Z and Z.Q.M conducted transport data analyses. All authors contributed to scientific discussions and read and commented on the manuscript. This project was supervised by Z.Q.M.

## Figure captions

**Figure 1 | Crystal structure, electronic and magnetotransport properties of $Sr_{1-y}Mn_{1-z}Sb_2$. a**, the crystal structure of $SrMnSb_2$. **b**, X-ray diffraction pattern of a typical single crystal on the (*h*00) plane. Inset: an optical image of a large single crystal. **c**, in-plane ($\rho_{xx}$) and out-of-plane ($\rho_{out}$) resistivity as a function of temperature; the $\rho_{out}/\rho_{in}$ ratio reaches ~609 at 2K, indicating a quasi-2D system. **d**, the out-of-plane magnetoresistivity, $\Delta\rho_{out}/\rho_{out} = [\rho_{out}(B)-\rho_{out}(0)]/\rho_{out}(0)$, as a function of magnetic field up to 31T at various temperatures. **e**, the low field out-of-plane magnetoresistivity. The Shubnikov-de Haas (SdH) oscillations extend down to 3T at low temperatures. **f**, the in-plane longitudinal ($\rho_{xx}$, upper panel) and transverse/Hall ($\rho_{xy}$, lower panel) resistivity, as a function of magnetic field at various temperatures. Both $\rho_{xx}$ and $\rho_{xy}$ exhibit strong SdH oscillations. The magnetic field was applied along the out-of-plane direction in the above magnetotransport measurements

**Figure 2 | Characteristics of relativistic fermions in $Sr_{1-y}Mn_{1-z}Sb_2$. a**, FFT spectra of $\Delta\rho_{out}(B)$ at various temperatures. $F_{2\alpha}$, $F_{3\alpha}$, $F_{4\alpha}$ and $F_{5\alpha}$ represent harmonic peaks. The inset shows the temperature dependences of the FFT amplitudes normalized by $\rho_{out}(B=0)$. The solid curves are the fits to the Lifshitz-Kosevich (LK) formula from 1.6 to 50K (black) and from 7 to 50K (red curve). **b,** the Landau Level (LL) index fan diagram derived from the SdH oscillations. The integer LL indices are defined from the minimum positions of conductivity $\sigma_{xx}$[45,46], obtained from $\rho_{xx}$ and $\rho_{xy}$ (Fig. 1f) using $\sigma_{xx} = \rho_{xx}/(\rho_{xx}^2+\rho_{xy}^2)$, as shown in the inset. The intercept $n_0$ = 0.46 on the LL index *n*-axis is consistent with a π Berry phase accumulated along the cyclotron orbit. **f**, Hall mobility $\mu_h$ as a function of temperature. Inset, the temperature dependence of Hall coefficient.

**Figure 3 | Coupling between magnetism and quantum transport properties in $Sr_{1-y}Mn_{1-z}Sb_2$. a**, field sweeps of magnetization at 5K for type A (showing significant FM behavior), B (showing relatively weak FM behavior), and C samples (displaying very weak FM behavior). All these samples exhibit de Haas–van Alphen oscillations with the amplitude of ~0.001-0.002$\mu_B$/Mn (see Fig. S6d for the dHvA oscillations for type A sample). **b**, SdH oscillations of $\rho_{xx}$ (upper panel) and $\rho_{out}$ (lower panel) of the type A, B, and C samples, with the magnetic field applied along the out-of-plane direction. The shift of the oscillation pattern is due to slight (< 10°) misalignment of magnetic field. **c**, temperature dependence of $\rho_{xx}$ (upper panel) and $\rho_{out}$ (lower panel) of the type A, B, and C samples. **d,** the relative SdH oscillation amplitude near 12.5T (upper panel) and the SdH/dHvA oscillation frequency (lower panel) as a function of the FM saturation moment $M_s$ for type A, B, and C samples. A general trend of increased oscillation amplitude and decreased oscillation frequency with increasing $M_s$ can be observed, as denoted by the dashed lines. The systematic variation of $F$ cannot be caused by the sample misalignment (< 10°) under magnetic field, as evidenced in the angular dependence of $F$ shown in Fig. S4c. **d,** the correlation between the relative SdH oscillation amplitude and the residual resistivity ratio RRR (= ($\rho$(300K)/$\rho$(2K), obtained from **c**).

**Figure 4 | Magnetism of $Sr_{1-y}Mn_{1-z}Sb_2$. a,** the temperature dependence of the diffraction intensity at (200) (normalized to 1 second) . The intensity at each temperature was obtained by counting 10 minutes. Inset, the FM magnetic structure of the Mn sub-lattice, viewed from the (101) direction. The Mn moments are aligned along the *b*-axis **b.** the magnetic structure of the FM state for $304\ K < T < 565\ K$. **c,** temperature dependence of the AFM order parameter (*i.e.*

the (001) magnetic peak intensity). The error bars are too tiny to be seen. Inset, the canted AFM magnetic structure of the Mn sub-lattice, viewed from the (101) direction. A net FM component due to moment canting is along the *b*-axis. **b.** the C-type AFM magnetic structure in the AFM state for $T < 304$ K.

# Figure 1

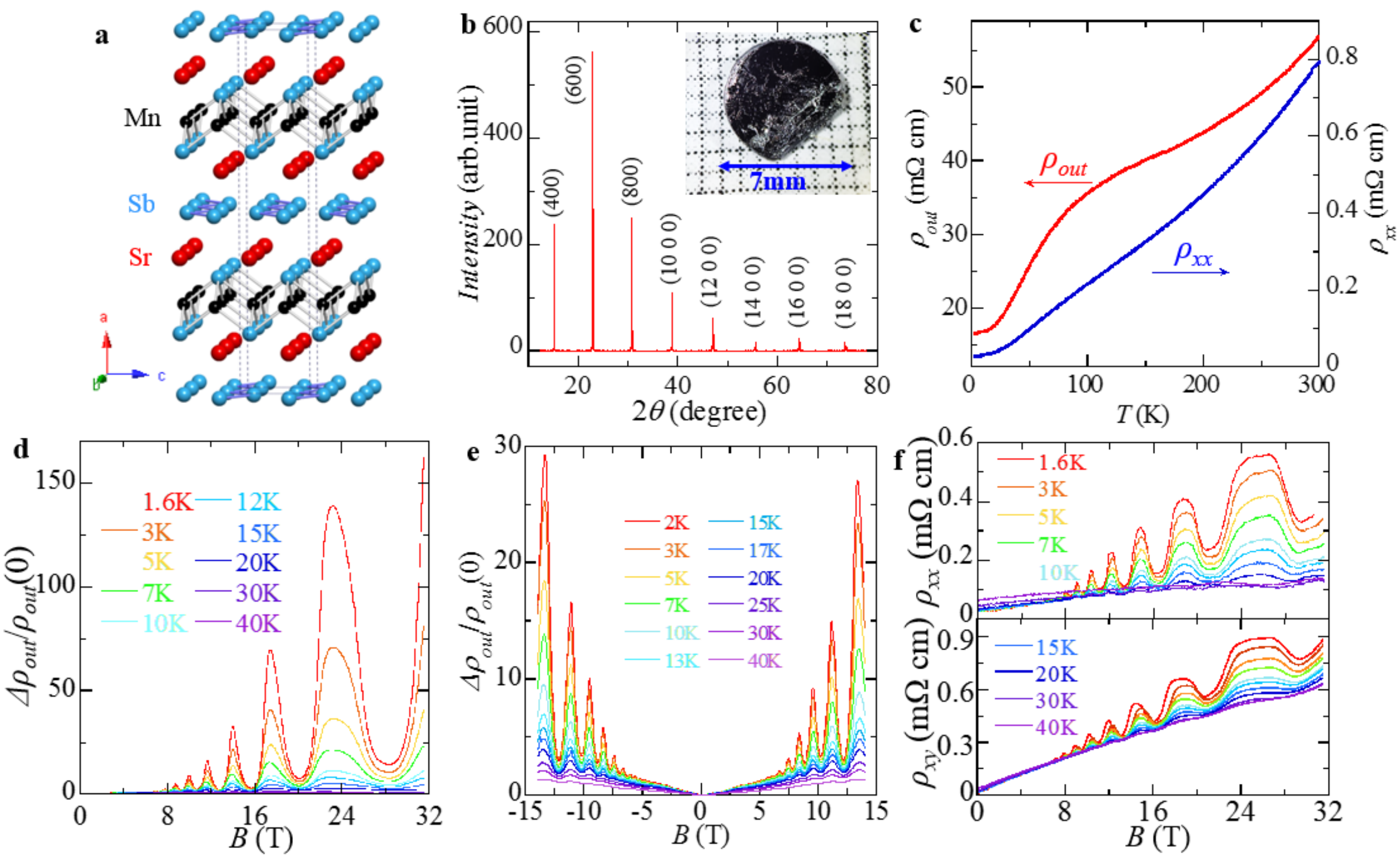

# Figure 2

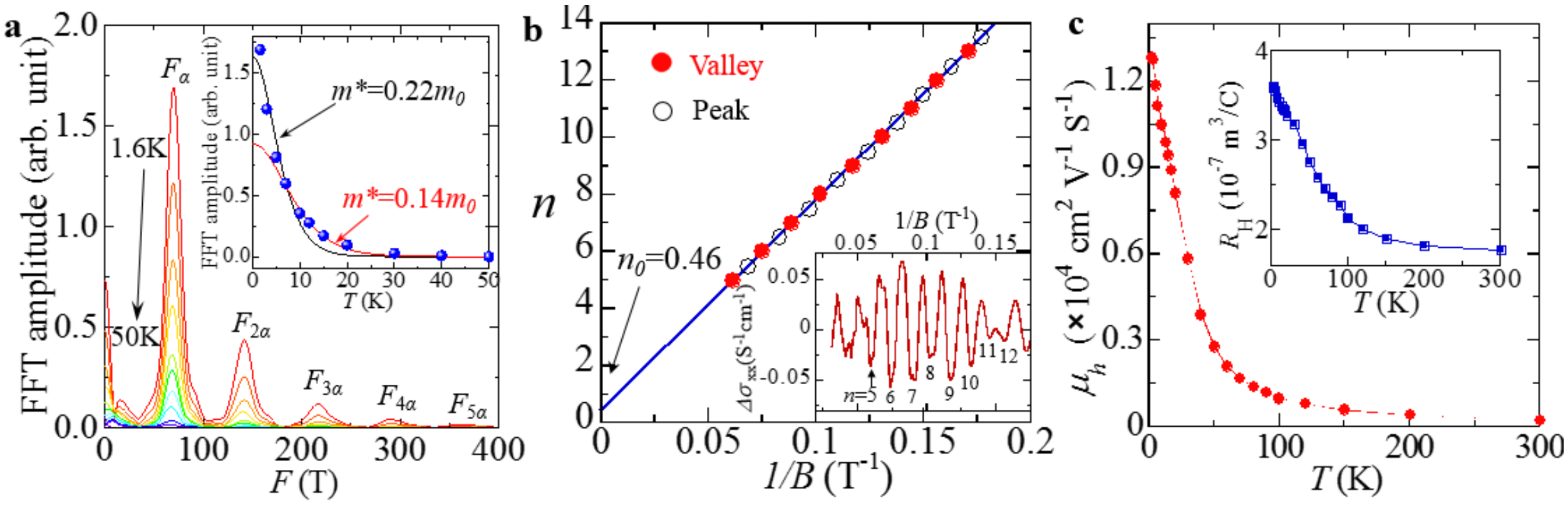

# Figure 3

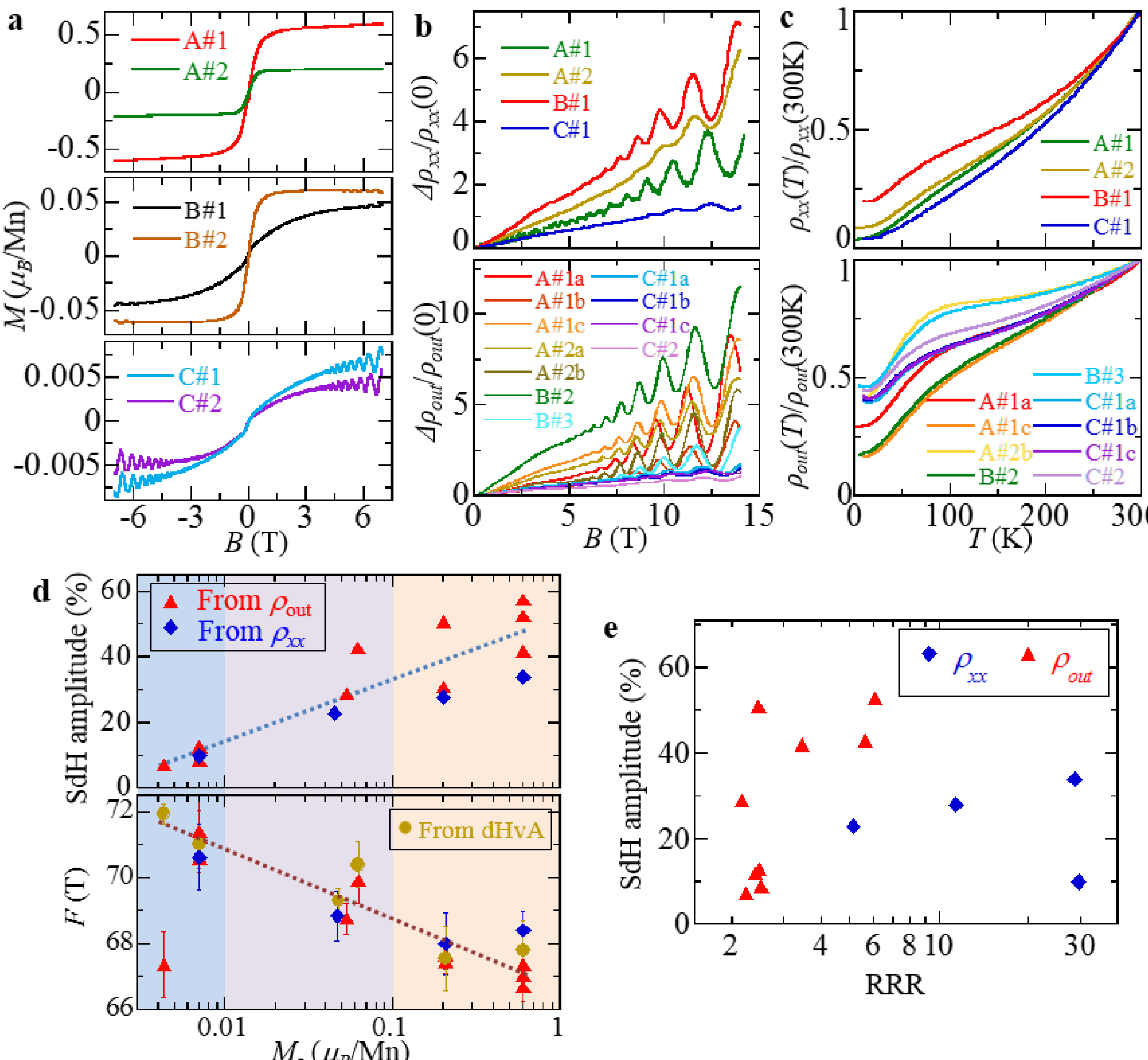

# Figure 4

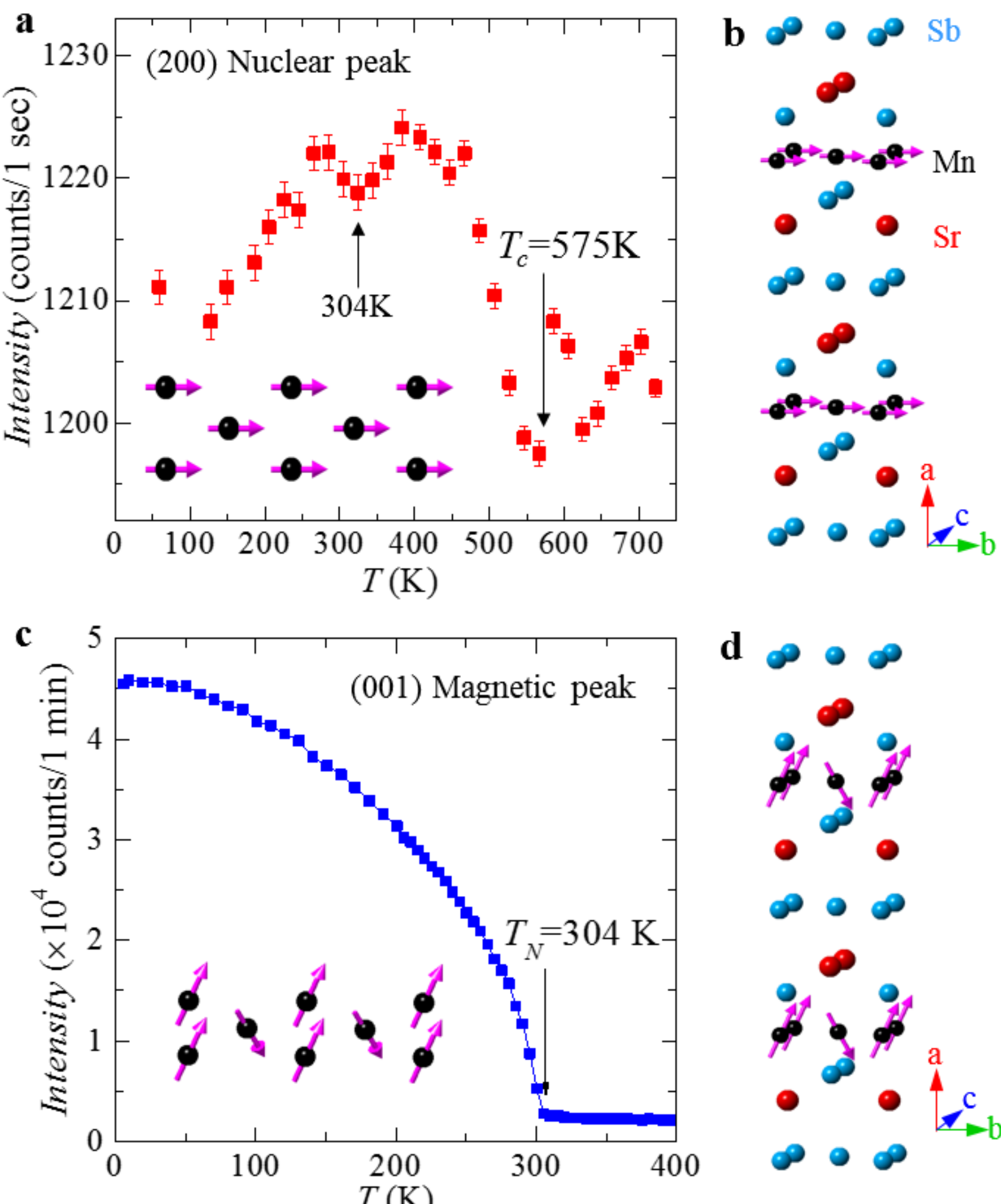

Supplementary Information for

# "Discovery of a magnetic topological semimetal $Sr_{1-y}Mn_{1-z}Sb_2$ ($y, z < 0.10$)"

## 1. Crystal structure of $Sr_{1-y}Mn_{1-z}Sb_2$

Extensive single crystal diffraction investigations were performed on two crystals of $Sr_{1-y}Mn_{1-z}Sb_2$ sample type A at HB-3A at HFIR, ORNL to determine the crystalline and magnetic structure. For the crystal structure, we collected the full data set at 315 K (above $T_N$=304 K) employing neutrons with a wavelength of 1.003 Å available via the silicon monochromator[1]. The lattice parameters and the observed reflections reveal an orthorhombic structure with the space group *Pnma*, similar to the structure of stoichiometric $SrMnSb_2$ as shown in Fig. 1. A comparison of the observed and calculated values of the squared structure factors from data taken at 315 K is shown in Fig. SI where $R_f$= 6.918% and $\chi^2$ = 3.125. The orthorhombic *Pnma* structure persists in our neutron data to lowest temperature of 5 K without any indication of a structural transition. We observed a twinning along the ***b/c*** axis in both crystals. The *b* and *c* lattice parameters were determined by aligning each structural domain of the twinned structure. The refined lattice parameters, atomic positions and reliability factors from our data taken at 315 K are summarized in Table S1.

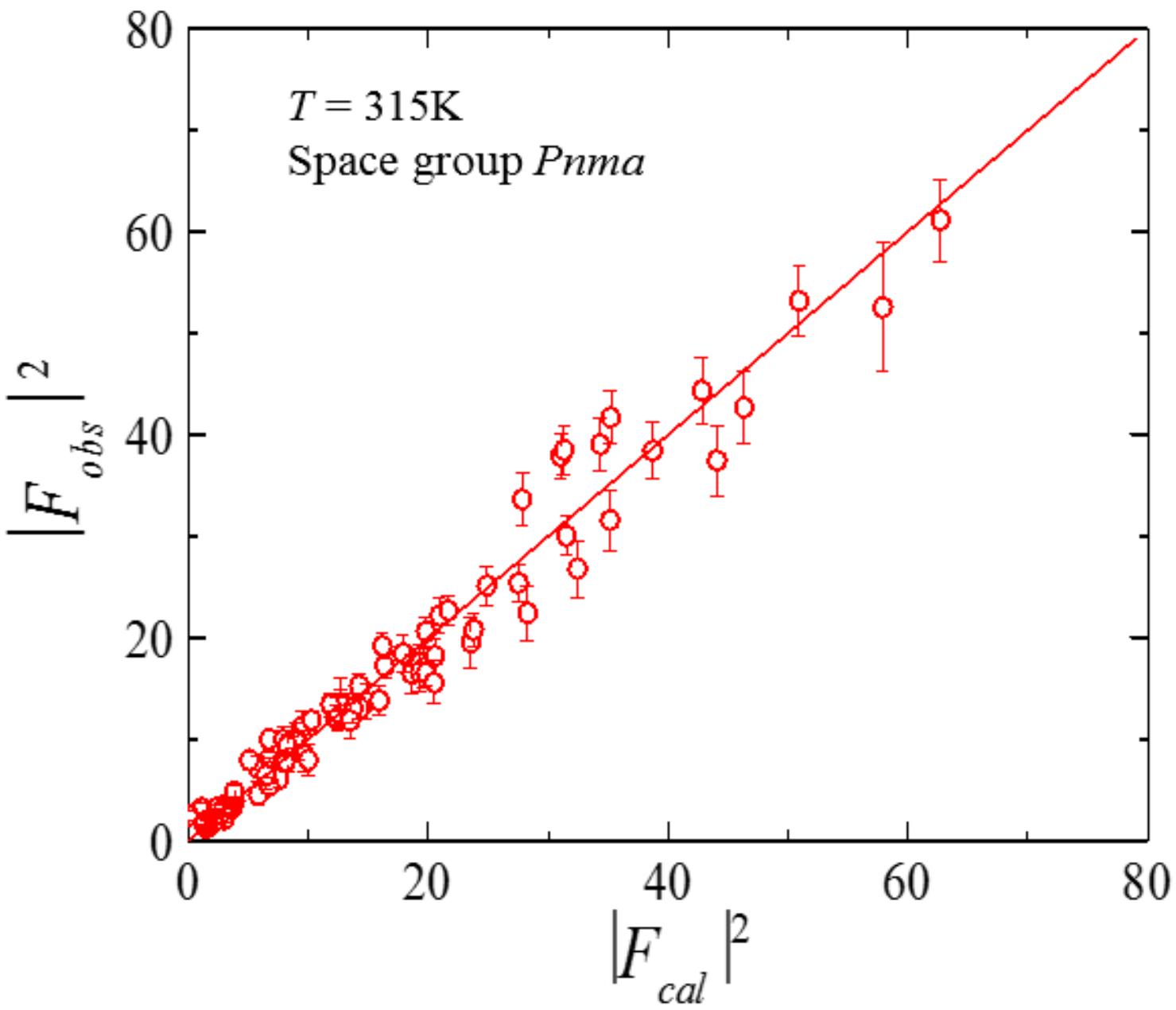

**Figure S1| Comparison of the observed and calculated values of the squared structure factors from our neutron diffraction data taken at 315 K**

**Table S1.** Atomic parameters determined from neutron diffraction from a single crystal of $Sr_{1-y}Mn_{1-z}Sb_2$ at 315 K. The space group, *Pnma* (No. 62) has been identified with lattice parameters, $\boldsymbol{a}$ = 23.011(4) Å, $\boldsymbol{b}$ = 4.384(5) Å, $\boldsymbol{c}$ = 4.434(7) Å, $\alpha = \beta = \gamma = 90^{\circ}$.

| | Wyckoff site | $x$ | $y$ | $z$ | Rf-factor | $\chi^2$ |
|---|---|---|---|---|---|---|
| Sr | 4c | 0.1139(9) | 0.25 | 0.7250(9) | | |
| Mn | 4c | 0.250(2) | 0.25 | 0.22509(2) | | |
| Sb1 | 4c | 0.0016(8) | 0.25 | 0.21830(9) | | |
| Sb2 | 4c | 0.323(3) | 0.25 | 0.724(1) | | |
| Reliable factors | | | | | 6.918% | 3.125 |

## 2. SdH oscillations and effective mass of $Sr_{1-y}Mn_{1-z}Sb_2$

The in-plane and out-of-plane oscillatory component, $\Delta\rho_{xx}$ (Fig. S1a) and $\Delta\rho_{out}$ (Fig. S1b), were obtained by subtracting the non-oscillatory background of the MR. The oscillation frequencies of the SdH oscillations determined by FFT of the $\Delta\rho_{xx}$ and $\Delta\rho_{out}$ (see Fig. 2a and Fig. S2a) are 69T and 67T for $\Delta\rho_{xx}$ and $\Delta\rho_{out}$, respectively.

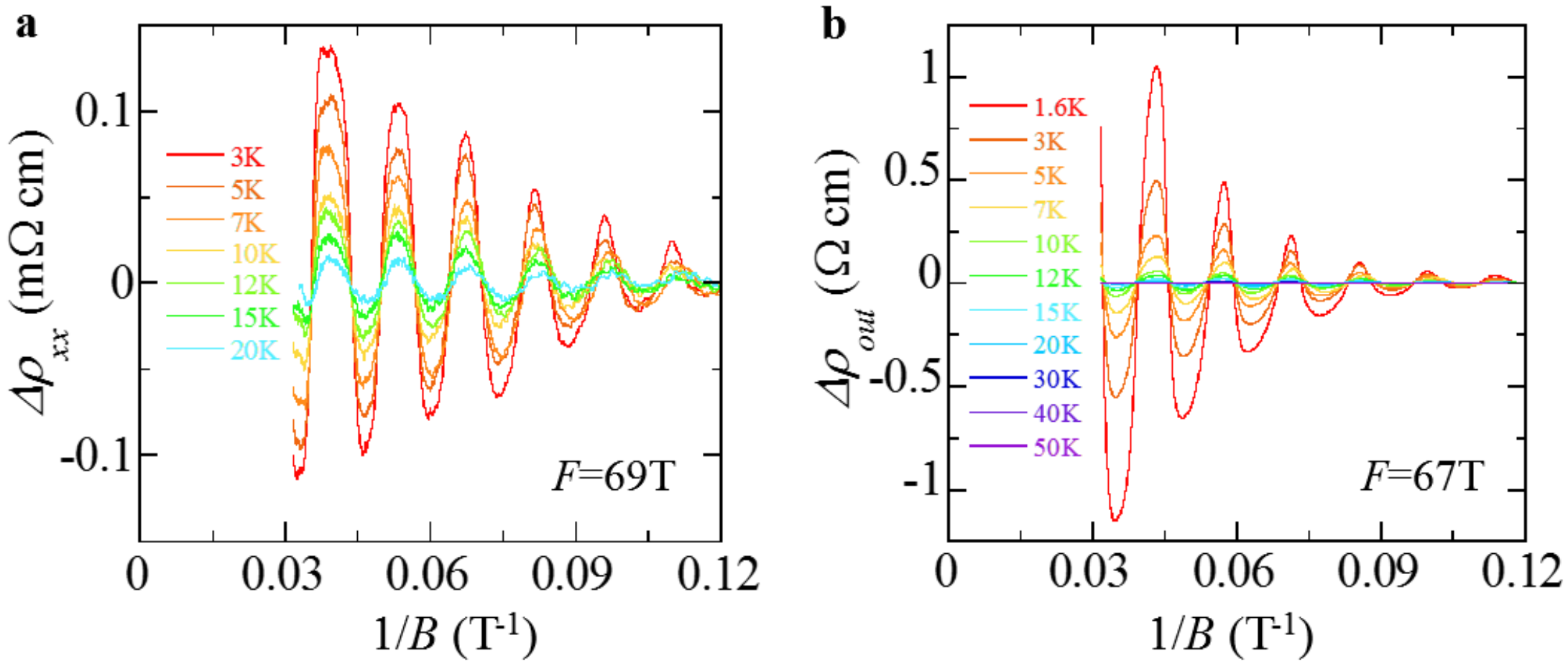

**Figure S2| The in-plane and out-of-plane oscillatory component of a) $\Delta\rho_{xx}$ and b) $\Delta\rho_{out}$**

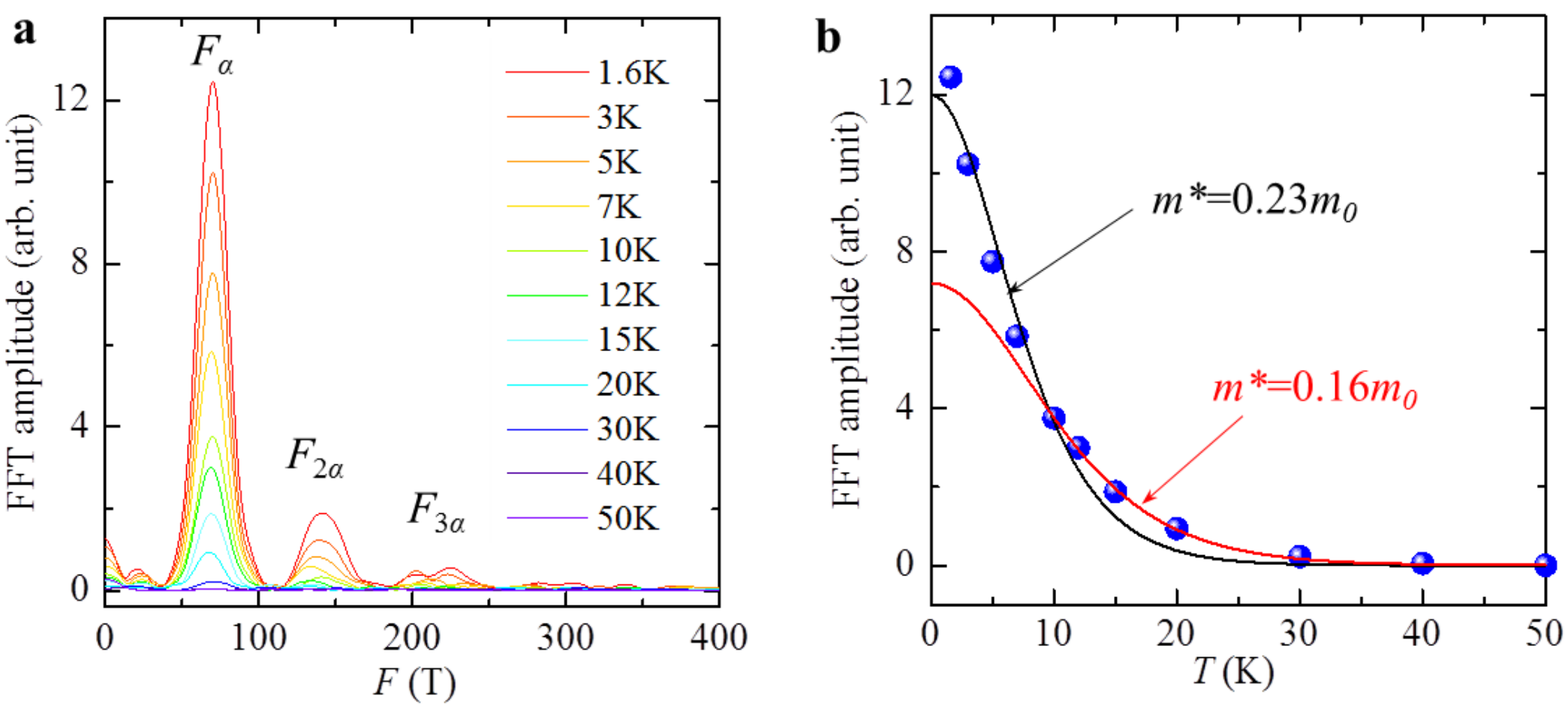

**Figure S3 | Determination of effective mass from in-plane resistivity oscillation. a,** FFT spectra of $\rho_{xx}$(B) at various temperatures. **b,** the temperature dependence of the FFT amplitude for $\rho_{xx}$. The solid curves are the fits to the Lifshitz-Kosevich (LK) formula from 1.6 to 50K (black) and from 7 to 50K (red curve).

As discussed in the main text, the satisfactory fitting of the FFT amplitudes for the temperature damping of oscillation amplitudes using the Lifshitz-Kosevich (LK) formula can only be obtained for high temperature data (7-50K). Similar situation also occurs in the in-plane resistivity sample. From the FFT spectra of $\rho_{xx}$ shown in Fig. S3a, we can obtain the temperature dependence of the FFT amplitude as shown in Fig. S3b. According to the LK formula, the thermal damping of FFT amplitude for $\Delta\rho/\rho_0$ is proportional to $\alpha T/\sinh(\alpha T)$, where $\alpha = (2\pi^2 k_B m^*)/(\hbar e \overline{B})$ with $1/\overline{B}$ being the average inverse field. However, as shown in Fig. S3, the LK formula barely fits the temperature dependence of FFT amplitude for the full temperature range from 1.6 to 50K, with the effective mass being determined to be 0.23$m_0$. Satisfactory fitting, however, can be obtained only when low temperature data points (<10 K) are not taken into account, which yields an effective mass of ~ 0.16$m_0$.

The deviation of FFT amplitude from LK formula is due to the quick enhancement of FFT amplitude at low temperature (<7 K). Mathematically, LK formula indicates a saturation of oscillation amplitude when approaching the zero temperature limit. However, the amplitude of resistivity oscillations (Figs. S2a and S2b) and the corresponding FFT amplitude (Figs. 2a and S2a) in our $Sr_{1-y}Mn_{1-z}Sb_2$ samples increase rapidly up on cooling, without any signature of saturation down to 1.6K. The origin of such unusual behavior is yet to be investigated.

## 3. Determination of Berry phase from conductivity oscillation

The Berry phase accumulated from the cyclotron motion of carriers can be determined from the LL index fan diagram, *i.e.,* the filling factor $n$ vs. inverse index field $1/B$. To obtain the correct Berry phase, the integer filling factor should be assigned by the magnetic field where the Fermi level lies in between two LLs [2,3], which corresponds to the minimum in conductivity oscillation.

In the practical experiments, the directly measured quantity is resistivity $\rho$, which is associated with conductivity by $\sigma_{xx} = \rho_{xx}/(\rho_{xx}^2+\rho_{xy}^2)$. Considering that resistivity $\rho$ oscillates around an average background value $<\rho>$, $\rho = <\rho> + \Delta\rho$ where $\Delta\rho$ is the oscillation component. The oscillation component of conductivity $\Delta\sigma_{xx}$ is determined by both average backgrounds and the oscillation components of $\rho_{xx}$ and $\rho_{xy}$. In the extreme case of $\langle \rho_{xx} \rangle << \langle \rho_{xy} \rangle$ and $\Delta\rho_{xx} << \Delta\rho_{xy}$, $\Delta\sigma_{xx}$ is in phase with $\Delta\rho_{xx}$ and the integer filling factor can be assigned by the resistivity oscillation minimum. Otherwise the conductivity oscillation should be used to precisely determine the Berry phase. In the case of $Sr_{1-y}Mn_{1-z}Sb_2$, given that $\rho_{xx}$ and $\rho_{xy}$ are comparable (Fig. 1f), we have used the minimum of the conductivity oscillation obtained by $\sigma_{xx} = \rho_{xx}/(\rho_{xx}^2+\rho_{xy}^2)$ to define the integer LL index $n$ (Fig. 2b). A Berry phase factor of $0.92\pi$ was revealed from our established LL index fan diagram (Fig. 2b), which is consistent with the expectation for relativistic fermions for a quasi-2D system[44].

## 4. 2D Fermi surface of of $Sr_{1-y}Mn_{1-z}Sb_2$

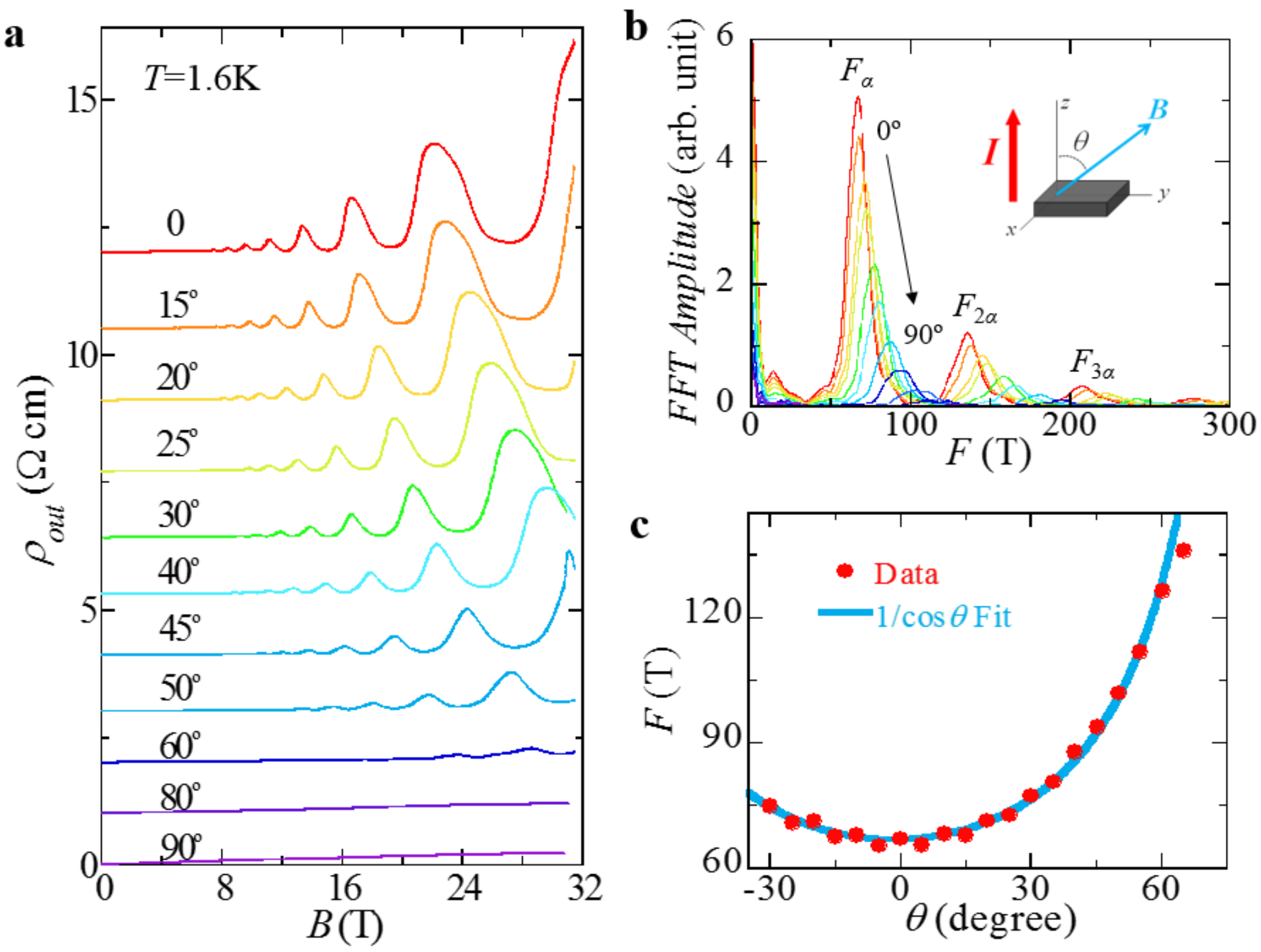

**Figure S4 | Determination of quasi-2D Fermi surface through the measurements of angular dependence of SdH oscillation frequency. a,** the out-of-plane resistivity $\rho_{out}$ of the A#1a sample as a function of magnetic field under different field orientation angles $\theta$ (the polar angle , measured from the out-of-plane direction, as shown in the inset to **b**). The data have been shifted for clarity except for the one taken $\theta = 90°$. **b**, FFT spectra of $\rho_{out}(B,\theta)$ shown in **a**. The inset illustrates the relative orientations of applied current and field for measurements shown in **a**. **c**, the $\theta$ dependence of the SdH oscillation frequency $F(\theta)$. $F(\theta)$ follows a $1/\cos\theta$ function as shown by the blue fitting curve, indicating that the FS responsible for the SdH oscillations is quasi-2D.

Like $SrMnBi_2$, $Sr_{1-y}Mn_{1-z}Sb_2$ should also be a quasi-2D system. Although there has been no ARPES studies on $Sr_{1-y}Mn_{1-z}Sb_2$, our measurements of SdH oscillation frequency $F$ as a function of field orientation angle demonstrate that the FS of this compound is indeed quasi-2D. Fig. S4a presents $\rho_{out}$ of the A#1a sample as a function of magnetic field applied along different orientation angles $\theta$ (see the inset to Fig. S4b) at 1.6K. The SdH oscillation peaks shift systematically with the increase of $\theta$. We performed FFT analyses for these data, as shown in Fig. S4b where three peaks are observed and two of them denoted by $F_{2\alpha}$ and $F_{3\alpha}$ are harmonic peaks. The angular dependence of the oscillation frequency $F(\theta)$ derived from FFT analysis is presented in Fig. S4c and can be fitted to a $1/\cos\theta$ function, suggesting the FS responsible SdH oscillations in $Sr_{1-y}Mn_{1-z}Sb_2$ is quasi-2D, consistent with the quasi-2D transport properties probed in the resistivity measurements (Fig. 1c).

The angular dependence of the oscillation frequency also indicates that the variation of quantum oscillation frequency with ferromagnetic (FM) strength (shown in the bottom panel of Fig. 3d) cannot be attributed to the field misalignment (<10 ˚). Although we did observe shift of SdH oscillation pattern due to misalignment (Fig. S4a), the misalignment of <10 ˚ can only cause a frequency shift by about 1T, which can not explain the decrease of the oscillation frequency of ~5 T from type C to type A samples as seen in Fig. 3d.

## 5. Angle dependence of SdH oscillations for $Sr_{1-y}Mn_{1-z}Sb_2$

An important property of quasi-2D Dirac fermions in $SrMnBi_2$ is that it allows controlling the valley degree of freedom by magnetic field [4]. That is, the in-plane magnetic field rotation can modulate the contribution of each valley to the out-of-plane conduction. Thus, the in-plane field rotation results in a four-fold symmetry in the out-of-plane conductivity as a function of the in-plane field orientation angle, $\sigma_{out}(\phi)$. For $SrMnBi_2$, the four-fold symmetry of $\sigma_{out}(\phi)$ can be quantitatively described by a function which expresses the total out-of-plane conductivity as the sum of conductivity contributed by each FS[4]. From this fit, an anisotropy factor $r$ of the effective mass can be evaluated. Given that $Sr_{1-y}Mn_{1-z}Sb_2$ has a similar quasi-2D

FS as $SrMnBi_2$ has, we naturally expect to observe a similar magnetic valley control in $Sr_{1-y}Mn_{1-z}Sb_2$. In order to verify this, we measured $\rho_{out}$ as a function of polar ($\theta$) and azimuthal ($\phi$) angles (see inset of Fig. S5d for the definitions of $\theta$ and $\phi$).

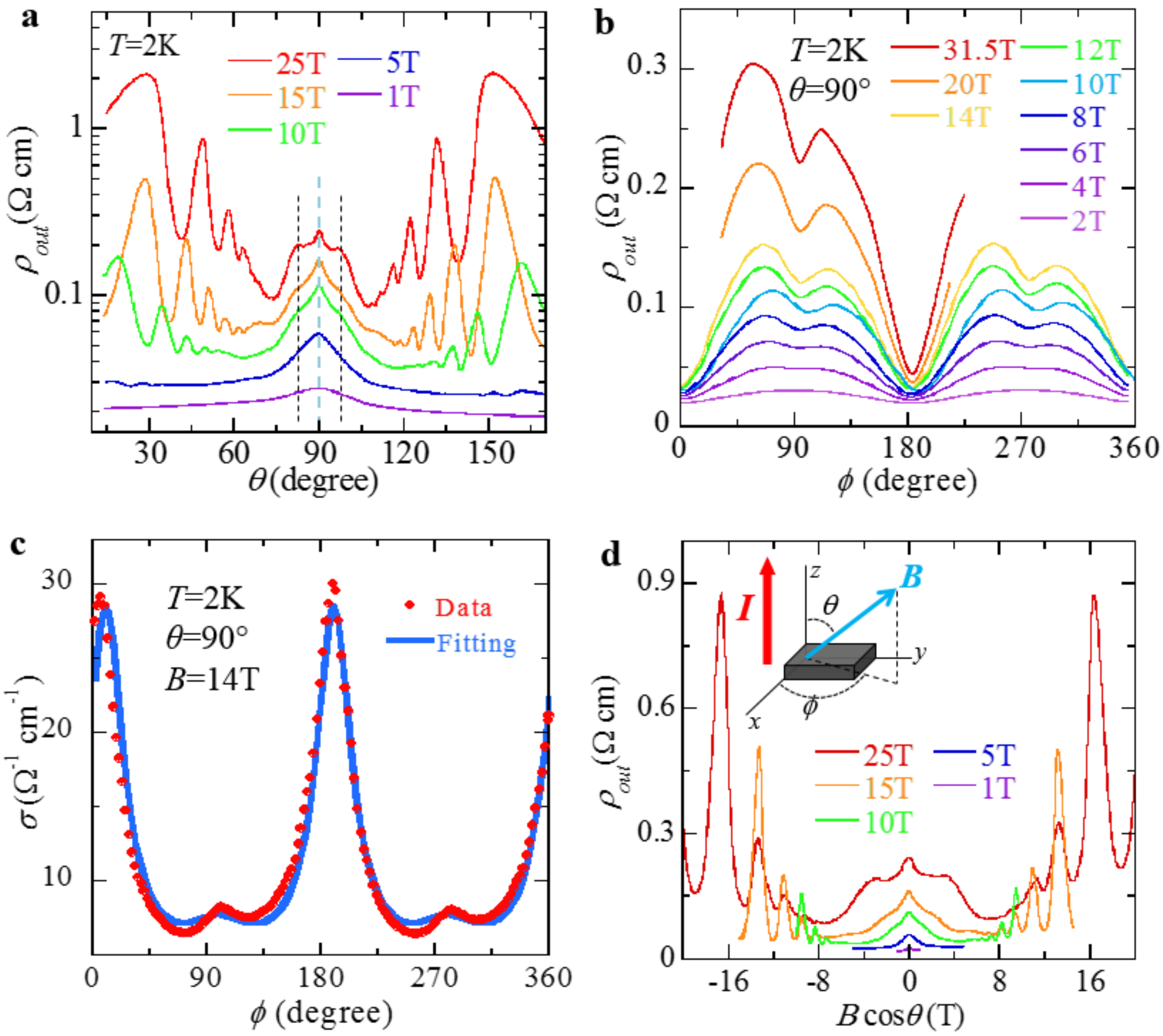

**Figure S5 | Valley polarized interlayer conduction of $Sr_{1-y}Mn_{1-z}Sb_2$. a**, the polar angle ($\theta$) sweep of $\rho_{out}$ of the A#1a sample under various magnetic fields at 2K. The peaks denoted the black dashed lines are the Yamaji peaks due to a geometric effect [4] and the central peak at $\theta$ = 90° is the coherent peak which can be modulated by the in-plane field rotation as seen in **b**. Other peaks seen in the $\theta$ sweeps are due to SdH oscillations. **b**, the azimuthal angle ($\phi$) sweeps of $\rho_{out}$ of the A#1a sample at $\theta$ = 90° under various fields at 2K. **c**, the fit of the out-of-plane conductivity $\sigma(\phi, \theta$=90°) measured under 14T at 2K to Eq. (1). **d**, angular dependent of SdH oscillations. Inset: the schematic of the current and field orientations relative to the sample. $\theta$, the

polar angle measured from the $z$-axis along which the current is applied; $\phi$, the azimuthal angle measured from the $x$-axis on the $xy$ plane.

As shown in Fig. S5a, $\rho_{out}(\theta)$ of the A#1a sample shows similar behavior as that seen in $SrMnBi_2$ [4]: We observed oscillation peaks occurring in the ranges $\theta < 74^o$ and $\theta > 106°$. As shown in Fig. S5d, these peaks follows the same oscillation pattern with respect to the perpendicular field component, $B_z$ (=$B\cos\theta$), indicating the origin of SdH oscillations in 2D system. In addition to SdH oscillation peaks, $\rho_{out}(\theta)$ also displays Yamaji peaks (denoted by the black dashed lines) due to a geometric effect and a coherent peak at $\theta$=90° [4]. As seen in $SrMnBi_2$, we find the coherence peak is also modulated by $\phi$ (see Fig. S4b) when $\theta = 90°$, indicating the presence of valley polarized interlayer conduction. However, in contrast with the 4-fold symmetry in $SrMnBi_2$ [4], we observed an anisotropy which can be viewed as a superposition of two-fold anisotropies with different amplitudes and a phase difference of $\pi/2$. This should be associated with the orthorhombic distortion in $Sr_{1-y}Mn_{1-z}Sb_2$.

Following the approach of quantitative analyses for $\sigma_{out}(\phi)$ in $SrMnBi_2$, we fitted our $\sigma_{out}(\phi)$ data collected at 2K under a field of 14T (Fig. S5c) to a modified function similar to eq. (1) in ref. [4]:

$$\sigma_c(\phi) = \sum_{n=1}^{4} \sigma_{\alpha,n}(\phi) + \sigma_\beta = \frac{2\sigma_{2D,1}}{1 + r_1 \cos^2(\phi)} + \frac{2\sigma_{2D,2}}{1 + r_2 \cos^2(\phi + \pi/2)} + \sigma_{3D} \qquad (1),$$

where the total conductivity is expressed as the sum of conductivity contributed by each FS, $\sigma_{2D}$ and $\sigma_{3D}$ represent the relative contribution of the $\alpha$ and $\beta$ FS respectively. Orthorhombic distortion has been taken into account in eq. (1); $r_1$ and $r_2$ represent $m^*$ anisotropic factors of two $\alpha$-FSs rotated by 90° in the BZ. As seen in Fig. S5c, our data can be approximately fitted by eq.

(1); the small deviation between the fitted curve and the experimental data near the extremal angle positions can be ascribed to a small Hall component involved in $\rho_{out}$, which was indeed reflected in the slight asymmetry between $\rho_{out}(+B)$ and $\rho_{out}(-B)$. This fit yields $(2\sigma_{2D,1}+2\sigma_{2D,2})/\sigma_{3D} \sim 9$, much greater than the value of $4\sigma_{2D}/\sigma_{3D}$ (~0.08) in $SrMnBi_2$ [4]; this implies that the $\beta$-FS in $Sr_{1-y}Mn_{1-z}Sb_2$ has minor contribution to transport or even does not exist, in a good agreement with our observation of significantly large electronic anisotropy and strong SdH oscillations. $r_1$ and $r_2$ obtained from the fit are ~3 and ~11 respectively, much less than the $r$ value (~50 at 2K and 14T) of $SrMnBi_2$, implying the Dirac cone in $Sr_{1-y}Mn_{1-z}Sb_2$ should be much less anisotropic.

## 6. Magnetism of $Sr_{1-y}Mn_{1-z}Sb_2$

**Figure S6 | Magnetic properties of $Sr_{1-y}Mn_{1-z}Sb_2$. a-c,** Temperature dependence of the out-of-plane magnetization at of **a**) type A, **b**) type B, and **c**) type C samples of $Sr_{1-y}Mn_{1-z}Sb_2$. **d-f,** the field dependences of the isothermal out-of-plane magnetization of the identical **d**) type A, **e**) type

B, and **f**) type C samples. Inset to **d)**, zoomed- in data of the out-of-plane magnetization at higher fields at 5K. dHvA oscillations can be observed.

As shown in Figs. S6a and S6d, both of our temperature- and field-dependence magnetization measurements have revealed significant ferromagnetic behavior for type A samples, with the out-of-plane saturation moment $M_s$ being ~ 0.4-0.6 $\mu_B$/Mn for $T$= 5-400 K. Such FM behavior is consistent with the results of our neutron scattering measurements on a type A sample with $M_s$ ~ 0.2 $\mu_B$/Mn, which revealed a FM order in 304K<$T$<575K and a canted AFM order with a FM component below 304K as stated in the text.

Different from the neutron scattering measurements, the FM-to-AFM transition near 304K (Fig. 4c) was not observed in our magnetization measurements on type A samples (Fig. S6a), which should be due to the strong FM component. Indeed, when FM component is weakened by an order of magnitude (i.e., type B samples), we have observed a decrease of magnetization below ~304K, a typical behavior of antiferromagnetism (Fig. S6b). At low temperatures, we observed a significant magnetization upturn, which should be attributed to the development of FM component due to canted AFM moments. Further reducing FM strength (i.e. type C samples) leads to more remarkable AFM signature, as well as suppressed low temperature magnetization upturn (Fig. S6c).

In the isothermal magnetization measurements, we have observed clear dHvA oscillations at low temperatures ($T$=5K) of type B and C samples. In fact, dHvA oscillation also occurs in type A samples, but only visible in the zoomed-in data due to the very strong background from the large FM component, as seen in the inset of Fig. S6d.

## 7. Magnetic structure of $Sr_{1-y}Mn_{1-z}Sb_2$

To investigate the magnetic structure, a full data set was collected at 5 K in addition to monitoring the temperature dependence of the (200) and (001) Bragg reflections in the temperature range of 5-700 K using the neutrons with a wavelength of 1.542 Å. We note that the incident beam includes ~1.4% λ/2 contamination from the Si-220 monochromator in high resolution mode (bending 150) [1]. All of the magnetic reflections can be indexed with the *Pnma* unit cell with a commensurate magnetic propagation vector (***k*** = 0) and all of the magnetic reflections above $T_N$=304 K coincide with structural Bragg peaks. A FM order was indicated by a rapid increase of the (200) peak intensity below ≈565 K (see Fig. 4a). The SARAH representational analysis program[5] was employed to determine the symmetry of the allowed magnetic structures. We summarize the basis vectors of the allowed magnetic structures in Table S2 where it is clear that only the Γ5 symmetry is true FM solution indicative of moment along *b* axis. The intensity increases at (200) below ≈565 K verified that the Γ5 magnetic structure for temperatures between 304 and 565 K since magnetic scattering of neutron measures the component of the magnetic moment that are perpendicular to the scattering vector. We determined the size of the magnetic moment based upon the rapid decrease in intensity as the sample is warmed above 565 K. To compare with magnetization data which is limited to T <= 400 K, we calculate the magnetic moment at 400 K to be ~ 0.49(8) $\mu_B$/Mn. The FM structure is displayed in Fig. 4b.

**Table S2**. The symmetry-allowed basis vector [$m_x$, $m_y$, $m_z$] for the space group *Pnma* with ***k***=(0,0,0) in $Sr_{1-y}Mn_{1-z}Sb_2$ (y, z < 0.10). Mn1: (0.250, 0.25, 0.225), Mn2: (0.750, 0.25, 0.275), Mn3: (0.750, 0.75, 0.775), Mn4: (0.25, 0.75, 0.725)

| IR | Γ1 | Γ2 | Γ3 | Γ4 | Γ5 | Γ6 | Γ7 | Γ8 |
|---|---|---|---|---|---|---|---|---|
| Mn1 | [0 my 0] | [mx 0 mz] | [mx 0 mz] | [0 my 0] | [0 my 0] | [mx 0 mz] | [mx 0 mz] | [0 my 0] |
| Mn2 | [0 -my 0] | [mx 0 -mz] | [mx 0 -mz] | [0 -my 0] | [0 my 0] | [-mx 0 mz] | [-mx 0 mz] | [0 my 0] |
| Mn3 | [0 my 0] | [-mx 0 -mz] | [mx 0 mz] | [0 -my 0] | [0 my 0] | [-mx 0 -mz] | [mx 0 mz] | [0 -my 0] |
| Mn4 | [0 -my 0] | [-mx 0 mz] | [mx 0 -mz] | [0 my 0] | [0 my 0] | [mx 0 -mz] | [-mx 0 mz] | [0 -my 0] |

Upon cooling below $T_N$≈304 K new magnetic reflections emerge as shown in Fig. S7a demonstrates by displaying rocking-curve scans of one representative (001) magnetic reflection at 400 K and 50 K. The integrated intensities of all the nuclear and magnetic reflections can be best fitted employing the Γ2 magnetic structure with the fits yielding $\boldsymbol{M_a}$=3.77(9) $\mu_B$. The comparison between observed and calculated values of the squared structure factor is shown in Fig. S7b where we find a $R_f$ of 6.99% and $\chi^2$ of 4.42. Thus, below 304 K the Mn spins at atoms 1 and 2 (see the caption of table S2 for identification of the Mn positions in the unit cell) are parallel and these are antiparallel to the mangetic moments at Mn atoms 3 and 4, forming a nearest-neighbor antiparallel alignment in *bc* plane and the parallel alignment along *a* axis., i.e., *C*-type AFM order with moment along *a* axis. In addition to the C-type AFM order we observed only a partial reduction in intensity of the (200) Bragg reflection indicating the persistence of a ferromagnetic moment below 304 K. In addition to the C-type AFM order we observed only a partial reduction in intensity of the (200) Bragg reflection indicating the persistence of a ferromagnetic order along *b* axis below 304 K. In consideration of the existence of ferromagnetic order along the ***b*** axis consistent with magnetization measurements below $T_N$, a *C*-type AFM order with noncollinear magnetic moments such that they are canted toward the ***b*** axis is indicated. The combination of Γ2 and Γ5 magnetic structures in Table S2 describes these observations well. Note that the neutron refinement cannot resolve directly the FM component at 5 K due to its weakness. Nevertheless, the FM moment $\boldsymbol{m_y}$ is estimated to be ~0.27 (8) $\mu_B$ at 60 K based on the intensity variation below/above $T_N$, consistent with ~0.2 $\mu_B$ at a lower temperature of 5 K in the magnetization measurements. The canted *C*-type AFM structure is displayed in Fig. 4d. In summary, $Sr_{1-y}Mn_{1-z}Sb_2$ (y, z < 0.10) exhibits a PM-FM magnetic transition at ≈565 K with moment of 0.41(7) $\mu_B$ at 400 K, followed by a second magnetic transition at $T_N$≈304 K from collinear FM order to canted *C*-type AFM order with $M_a$ and $M_b$ of 3.77(9) $\mu_B$ and 0.2 $\mu_B$, respectively, at 5 K.

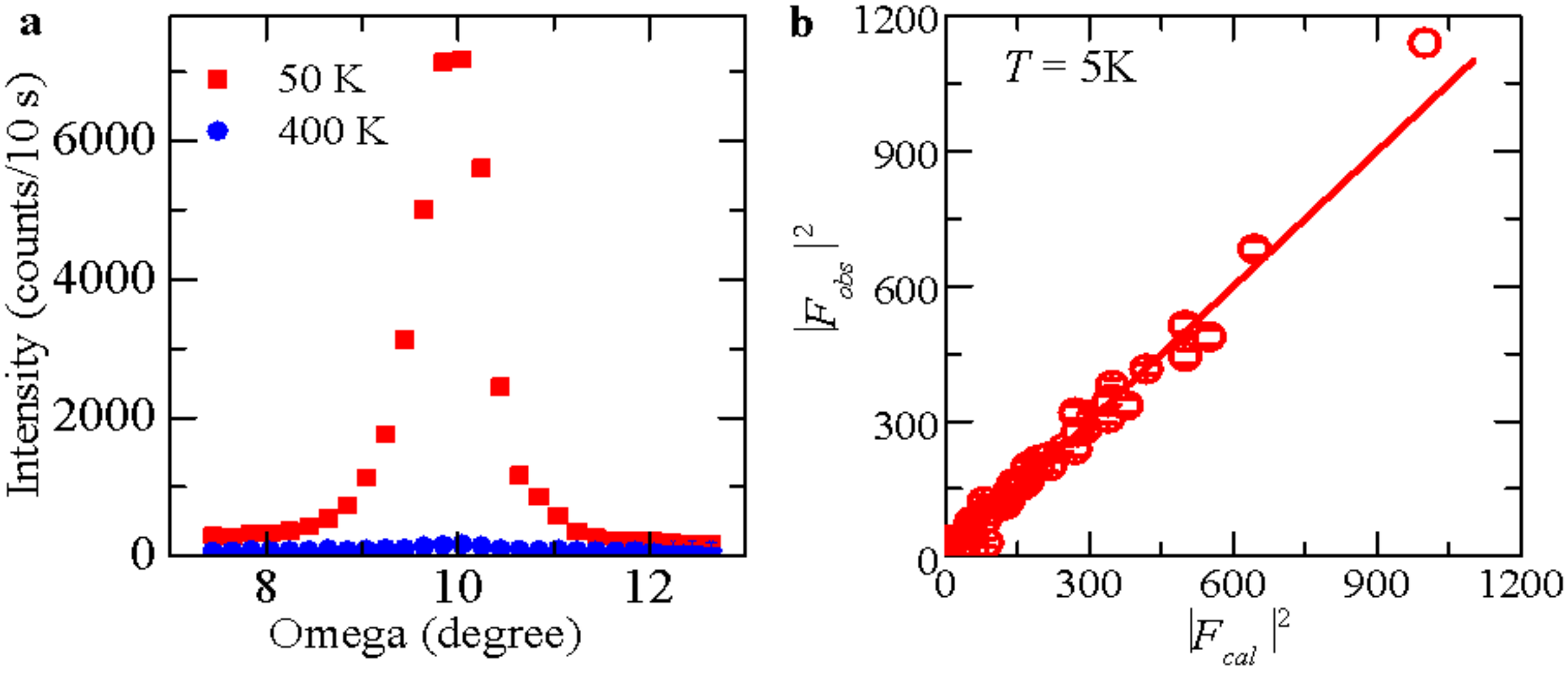

**Figure S7 | a**, Rocking-curve scans of the (001) magnetic reflection at 400 K and 50 K. The weak intensity observed at 400 K results from λ/2 contamination (λ=1.542 Å). The error is to tiny to be seen. **b**, Comparison of the observed and calculated squared structure factors for the Γ2 magnetic configuration at 5 K

## 8. Negative longitudinal magnetoresistance in $Sr_{1-y}Mn_{1-z}Sb_2$

In Weyl semimetals, the pair of Weyl nodes acts as source and drain of Berry flux, leading to a non-zero Berry curvature $\mathbf{\Omega}_p$ and causing an additional topological contribution to the Weyl fermions dynamics, which is proportional to the product of electric and magnetic field, *i.e.*, $\propto(\boldsymbol{E}\cdot\boldsymbol{B})\Omega_p$ [6]. As a consequence, non-orthogonal electric and magnetic field ($\boldsymbol{E}\cdot\boldsymbol{B}\neq0$) can lead to charge transfer between two Weyl nodes with opposite chirality. Such violation of the chiral charge conservation is known as the Adler-Bell-Jackiw anomaly or chiral anomaly [6-8], and results in negative MR that can take place in semiclassical regime [6,8].

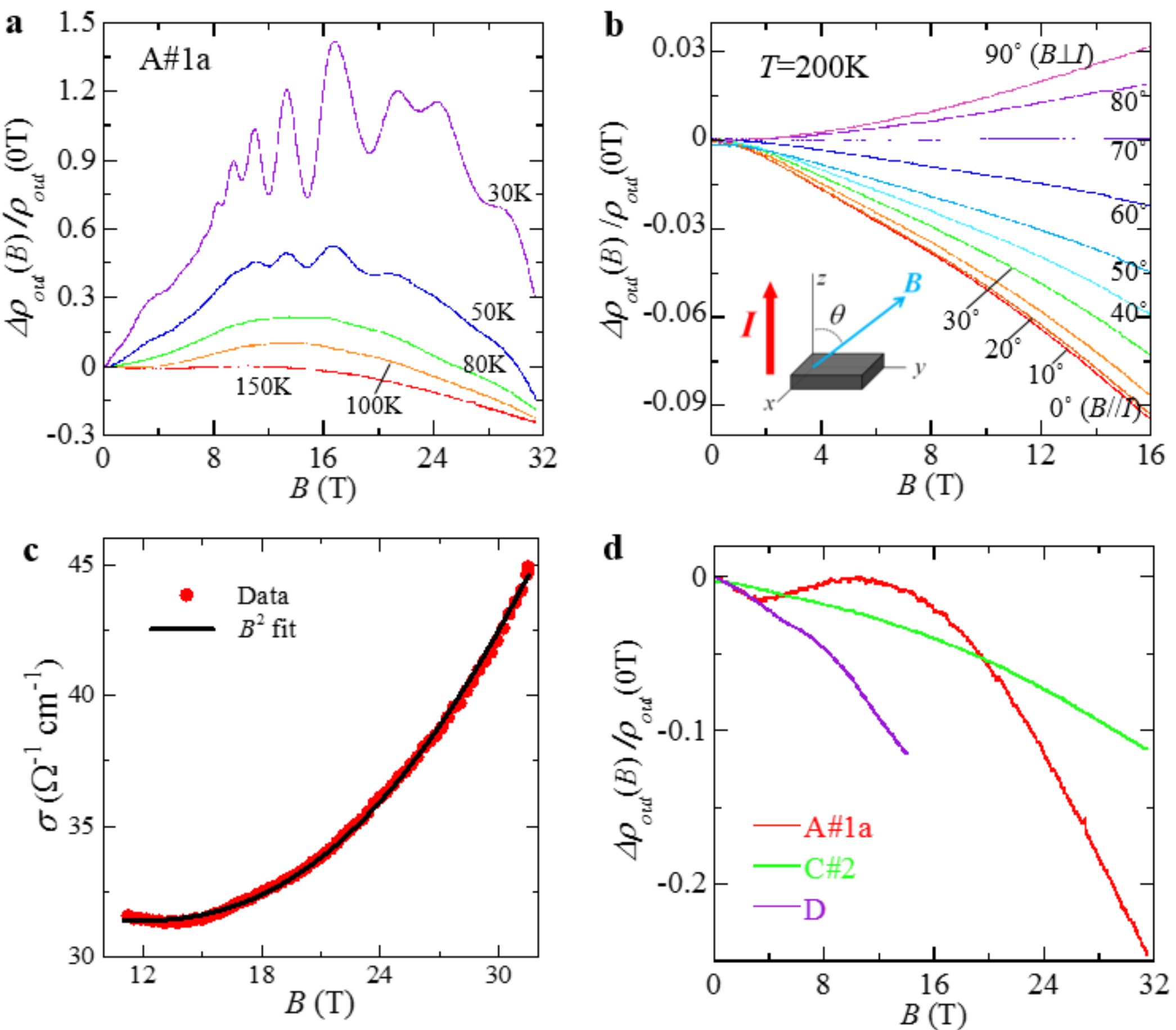

**Figure S8 | Negative longitudinal magnetoresistance in $Sr_{1-y}Mn_{1-z}Sb_2$. a,** The normalized longitudinal MR, $\Delta\rho_{out}(B)/\rho_{out}(0\text{T}) = [\rho_{out}(B) - \rho_{out}(B{=}0)]/\rho_{out}(B{=}0)$ for T= 30-150K for the identical A#1a sample used in the main text. **b,** Orientation dependence of negative MR at *T*=200K. The negative MR gradually weakens and turns to be positive when the deviation of magnetic field ***B*** from current ***I*** is larger than 70º. Inset: the schematic of the measurements. **c,** The $\boldsymbol{B}^2$ fit for the magnetoconductance $\sigma$ at 150K. **d,** the longitudinal MR for serval type A $Sr_{1-y}Mn_{1-z}Sb_2$ samples, collected at *T* =150K for sample A#1a and C#2 and 200K for sample D.

In the identical out-of-plane resistivity sample A#1a described in the main text (see Fig. 3), we have observed the negative MR when the field is orientated parallel to current direction (***B***//***I***). As shown in Fig. S8a, at low temperatures (*T*=30K), the MR is positive and increases with field at low fields. However, with further increasing field above 16T, the resistivity starts to drop, without signature of saturation up to 31T (the highest field we can achieve). Raising temperature gradually suppresses the low field positive MR, and the signature of negative MR become more significant (Fig. S8a). This can be understood in terms of the suppression of the positive classical orbital MR at high temperatures. More interestingly, the negative MR persists even up to 200K, reaching a magnitude of about -10% at 16T (Fig. S8b).

The evolution of MR with field orientation is also consistent with the scenario of chiral anomaly-induced negative MR. As shown in Fig. S8b, we observed a gradually weakening of the negative MR with rotating field away from the current direction, from -10% for $B//I$ to +3% for $B\perp I$. This is consistent with the observation in Weyl semimetals[9-15]. Furthermore, the field dependence of the negative MR (positive magnetoconductance) can be fitted to $B^2$ dependence (Fig. S6c), in good agreement with the predicted quadratic field dependence in semiclassical regime[6,8].

The negative longitudinal MR has been reproduced in another two out-of-plane resistivity samples, as shown in Fig. S8d, but none of the in-plane resistivity samples show negative MR. It is important to clarify whether such negative MR in our $Sr_{1-y}Mn_{1-z}Sb_2$ sample has a magnetic origin, given that the ferromagnetism is known to give rise to negative MR due to spin polarization. However, in terms of the magnetic properties of $Sr_{1-y}Mn_{1-z}Sb_2$, the negative MR is less likely to be associated with magnetism. First of all, the magnetic moment in $Sr_{1-y}Mn_{1-z}Sb_2$ saturates rapidly with the field is increased above 2T at various temperatures from 2 to 400K (Figs. 3a and S6b), which is inconsistent with the observation of the unsaturated negative MR up to 31T. Secondly, in the C#2 sample whose saturation moment is two orders of magnitude lower than that of the A#1a sample, we can still observe very significant negative MR which is about half of that of the A#1a sample, implying that the negative longitudinal MR is not strongly coupled with the ferromagnetism. However, it is not clear why negative longitudinal MR does not occur in all samples. It may be due to degraded sample quality. In fact, the negative longitudinal MR in A#1a sample disappears after one month though the sample is stored in the inert gas atmosphere and SdH oscillations are still present.